\documentclass{aa}  

\usepackage{graphicx}
\usepackage{ulem}
\usepackage{txfonts}
\usepackage{xcolor}
\usepackage[colorlinks=true, linkcolor=blue, citecolor=blue, filecolor=green, urlcolor=blue]{hyperref}

\begin{document}

   \title{Detection of a cyclotron line  in the Be X-ray pulsar IGR J06074$+$2205}

   \subtitle{}

   \author{Kinjal Roy,\thanks{kinjal@rrimail.rri.res.in}
          Rahul Sharma,
          Hemanth Manikantan,
          \and
          Biswajit Paul
          }

   \institute{Raman Research Institute, C. V. Raman Avenue, Sadashivanagar, Bengaluru - 560 080, India.\\
             }

   \date{}

 
  \abstract
   {IGR J0607.4+2205 is a transient Be X-ray binary discovered two decades ago. IGR J0607.4+2205 underwent an outburst in 2023 during which it was observed twice with \textit{NuSTAR}.}
   {The main goal of this work is to model the broadband X-ray spectrum of IGR J0607.4+2205 during the outburst and to study the variations of the spectral and timing features at different intensities.}
   {We extracted the light curve and spectrum of the source from the two \textit{NuSTAR} observations carried out during the recent outburst in the energy range of 3$-$78 keV. We used the epoch folding technique to find pulsation from the source and to study the changes in emission characteristics from the source with energy across an order of magnitude variation in source luminosity.}
   {IGR J0607.4+2205 shows pulsations with a period of $\sim$347.6 s during both the observations, with a pulse fraction of $\geq$50\%. The broadband spectrum of the source was modelled using a power-law continuum with a high-energy cutoff. During the first observation, a cyclotron absorption line at $\sim$51 keV was also present in the source with an optical depth of $\sim$1.3. However, no cyclotron line feature was detected in the second observation when the source was an order of magnitude fainter. Additionally, soft excess was detected in the second observation, which was modelled with a black body component emerging from close to the neutron star (NS).}
   {We report the first ever detection of a cyclotron line in the broadband spectrum of IGR J0607.4+2205 centred at 51$\pm$1 keV. The magnetic field strength of the NS is estimated to be $\sim$4$\times$$10^{12}$ G from the centroid energy of the absorption line. A significant change is observed in the pulse profile with luminosity during the decay of the outburst, indicating an associated change in the beaming pattern.}

   \keywords{pulsars: individual (IGR J0607.4+2205) --- X-rays: binaries --- X-rays: bursts}

   \titlerunning{Detection of a cyclotron line in IGR J06074$+$2205}
   \authorrunning{Roy et. al.} 
   \maketitle

%

\section{Introduction}

Transient X-ray binary pulsars (XRPs) with a Be-type main sequence companion (BeXRPs) are useful for studying accretion onto highly magnetised neutron stars (NSs). The BeXRPs are known to exhibit a wide range of luminosity, ranging from $10^{37-38}$ erg s$^{-1}$ during the peak of the outburst to $10^{32-34}$ erg s$^{-1}$ during the quiescence phase~\citep{Be_transient_Tsygankova_2017}. Many BeXRPs have been detected since the launch of the first X-ray telescopes~\citep{Fermi_GBM}. Pioneered by the highly successful \textit{RXTE}-ASM~\citep{RXTE_ASM}, the practice of initiating observations soon after the onset of an outburst has become standard. Recent advancements in sensitive, wide-viewing-angle X-ray instruments such as \textit{MAXI}/GSC~\citep{maxi_main_paper}, \textit{Swift}/BAT~\citep{SWIFT_BAT}, and \textit{Fermi}/GBM~\citep{Fermi_GBM} enable the prompt identification of transients entering an outburst. Subsequent follow-up observations with sensitive broadband instruments like \textit{NuSTAR}~\citep{NuSTAR_paper_FH} allow comprehensive studies of the source characteristics across a wide range of energy.

In highly magnetised (B $\ge$ $ 10^{12}$ G) high-mass X-ray binary (HMXB) systems, the accreted matter gets channelled along the magnetic field lines to the poles of the NS. A hot spot is produced in the accretion mound near the magnetic poles of the NS, producing soft X-rays that undergo inverse-Compton scattering by electrons in the accretion column giving rise to strong X-ray emissions~\citep{Becker_2007}. The presence of strong magnetic fields at the polar regions leads to the quantisation of electron energy levels ($E_n$) in accordance with the Landau levels~\citep{Meszaros_1992_book} given by $ E_n \sim 11.6 \times n \times B_{12}$ keV, where $B_{12}$ is the magnetic field strength in units of $10^{12}$ G and $n$=1,2,3,4,.. denotes the different levels. The cyclotron resonant scattering feature (CRSF) is the absorption feature in the spectra of pulsars due to interactions between photons and quantised electrons present in the line-forming region of the NS. CRSFs are the best diagnostic tool for studying the magnetic field strength of a NS. The CRSF is detected in the hard X-ray regime and is observed as absorption-like features~\citep{CRSF_review_Staubert_2019}.  

The source IGR J0607.4+2205 was discovered by the \textit{JEM-X} instrument on board \textit{INTEGRAL}~\citep{Detection_Chenevez_2004}. A follow-up \textit{Chandra} observation of the source was performed to localise the source~\citep{BeXRB_Tomsick_2006ATel}. Subsequent observations of the source using \textit{XMM-Newton}~\citep{XMM_Newton_pulsation_Reig_2018} revealed pulsations at a period of 347.2 s, establishing the compact object as a NS. The companion of IGR J0607.4+2205 was found to be a B0.5Ve star establishing the source as a BeXRP, and the distance to the source was estimated to be 4.5 kpc~\citep{Distance_Reig_2010}. The source IGR J0607.4+2205 underwent an outburst in the beginning of October 2023 with coherent pulsation detected at 2.6700(2) mHz from \textit{Fermi}/GBM~\citep{FermiGBM_Malacaria_2023ATel}. The source underwent three outbursts during 2022-2023 as observed with \textit{MAXI}/GSC, and based on the assumption that the outbursts were type-I outbursts, the orbital period was calculated to be either 80 days or an integral division of the same; that is, 80/n days, where n=2,3...~\citep{MAXI_P_orb_ATel_2023}. However, further observations are required to find the exact orbital period of IGR J0607.4+2205.

\textit{NuSTAR} subsequently made two target of opportunity (TOO) observations of the source after the detection of the outburst in 2023. We report the results obtained from the spectral and timing analysis of the two TOO observations of IGR J0607.4+2205. The primary focus of this work is on the broadband spectral characteristics of IGR J0607.4+2205 and the detection of a CRSF at $\sim$ 51 keV.


\section{Observations and data reduction}

\begin{figure}
\centering
\includegraphics[width=0.95\hsize]{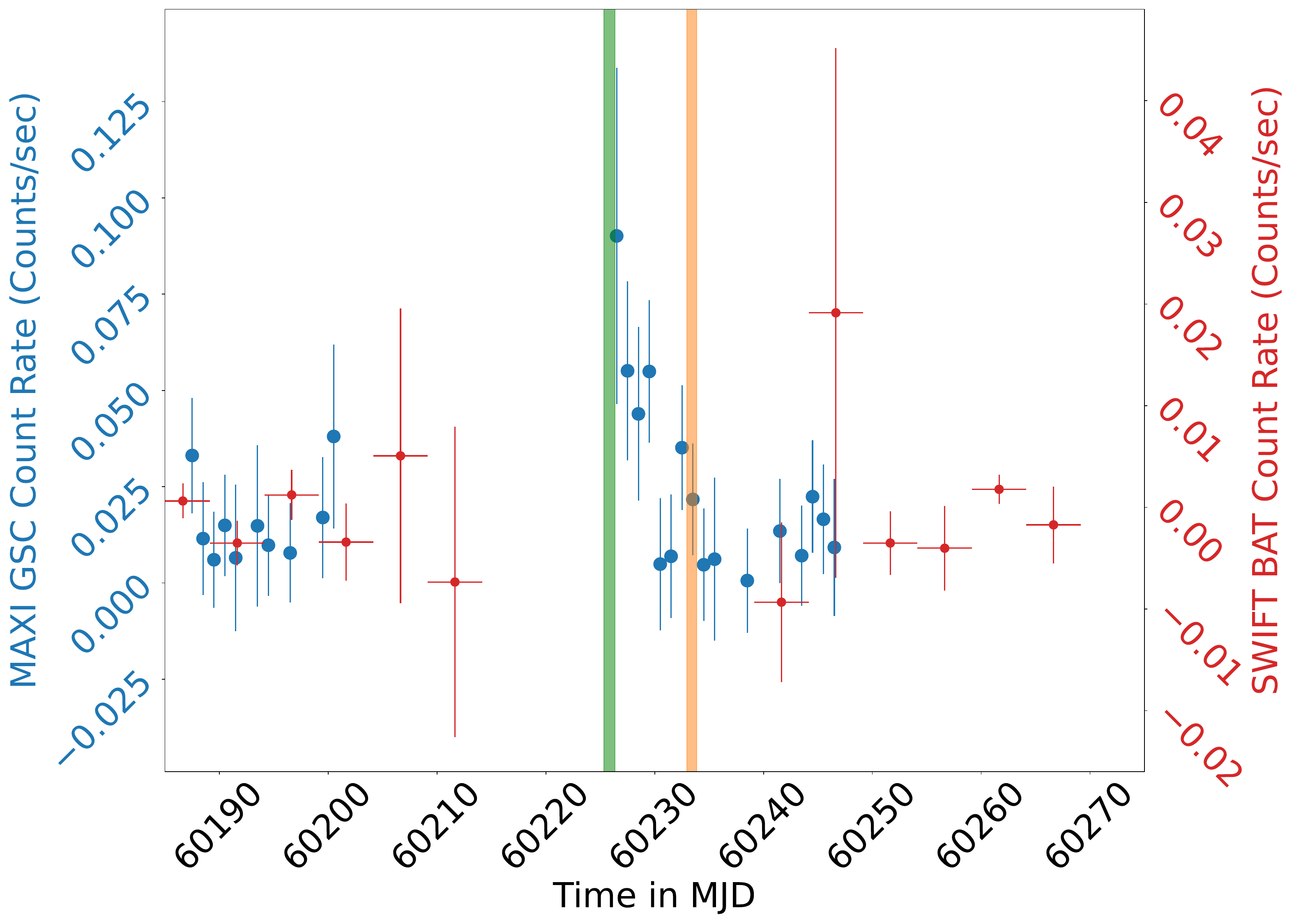}
    \caption{Long-term light curve from \textit{MAXI}/GSC (in blue) and \textit{Swift}/BAT (in red) plotted with a bin size of 1 day and 5 days, respectively. The green line denotes the epoch and duration of the first observation, while the second observation is shown in orange.}
    \label{Fig:2023 Outburst NuSTAR observation}
\end{figure}

The light curve for the October 2023 outburst was obtained from the \texttt{MAXI}\footnote{\url{http://maxi.riken.jp/mxondem/}} website in the 2$-$20 keV energy range. The light curve was also obtained from \textit{Swift}/BAT transient monitoring mission\footnote{\url{https://swift.gsfc.nasa.gov/results/transients/}}. The light curves from \textit{MAXI}/GSC and \textit{Swift}/BAT are plotted in Fig.~\ref{Fig:2023 Outburst NuSTAR observation} with a bin size of 1 day and 5 days, respectively.

The first of the two TOO observations was made on 8 October 2023 (MJD 60225; Obs ID: 90901331002; henceforth ObsID 02) with an effective exposure of $\sim$ 42 ks, while the second observation was made on 15 October 2023 (MJD 60232; Obs ID: 90901331004; henceforth ObsID 04) with an effective exposure of $\sim$ 40 ks. The two \textit{NuSTAR} observations are marked in Fig. \ref{Fig:2023 Outburst NuSTAR observation} along with the 2$-$20 keV \textit{MAXI}/GSC and 15$-$50 keV \textit{Swift}/BAT light curves of the source. 

The \textit{NuSTAR} data were reduced using \texttt{HEASOFT} version \texttt{6.30.1} along with the latest calibration files available via \texttt{CALDB} version \texttt{20231205}. The \texttt{nupipeline} version \texttt{0.4.9} was used to produce the clean event files and the \texttt{nuproduct} command was used to extract the light curves and spectrum of the source and background regions. A circle of  150 arcsec in radius around the source position was used to extract the level 2 source products and a circle of the same radius away from the source region was used to extract the background products for ObsID 02. For ObsID 04, the source and background region were respectively selected from a circle of  80 arcsec in radius around the source position and a circle of similar radius away from any known sources. Barycenter correction was performed using \texttt{barycorr} version \texttt{2.16}. All the spectral files were optimally binned~\citep{Optimal_binning_Kaastra_Bleeker_2016}. The broadband (3$-$78 keV) count rate of the source changed from 36 cts/s for ObsID 02 to 1.28 cts/s in the second observation.


\section{Analysis and results}

\subsection{Timing analysis}

\begin{figure}
    \centering
    \includegraphics[width=0.95\hsize]{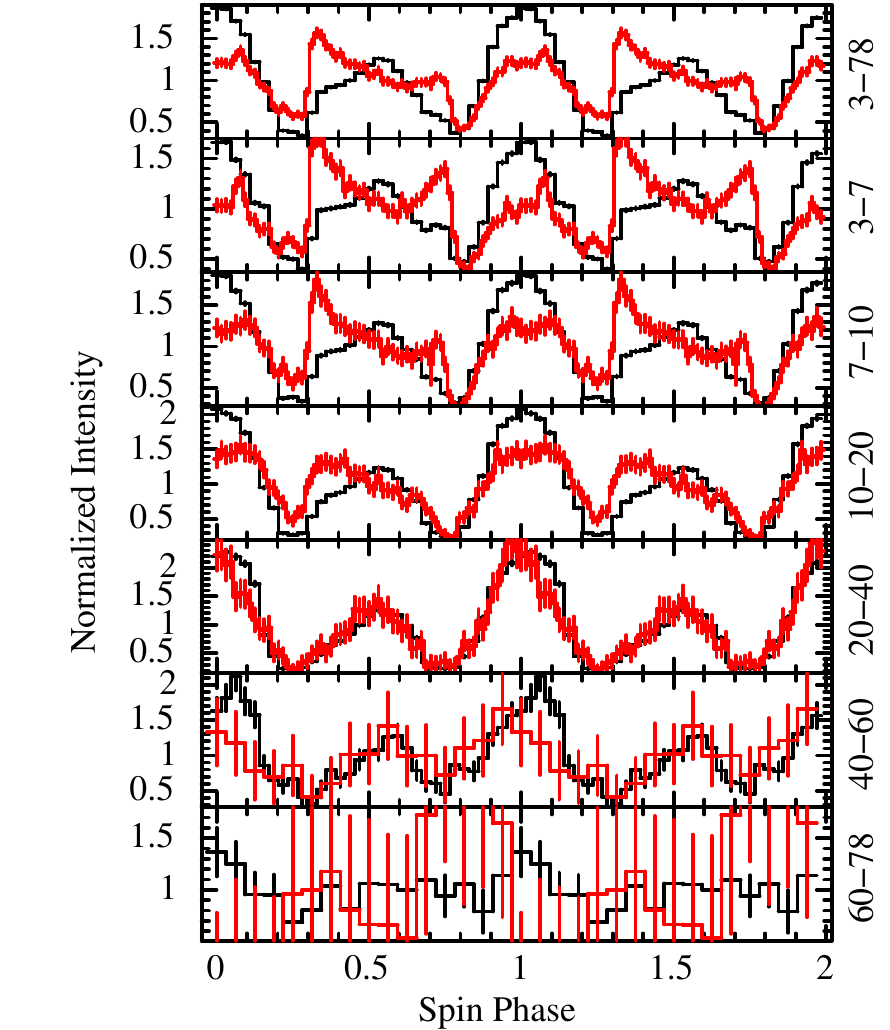}
    \caption{Normalised pulse profiles of combined FPMA and FPMB light curves from ObsID 02 (\textit{black}) and ObsID 04 (\textit{Red}). The top panel consists of the pulse profile in the energy range of 3$-$78 keV. The remaining panels show the energy-resolved profiles in the energy ranges of 3$-$7, 7$-$10, 10$-$20, 20$-$40, 40$-$60, and 60$-$78 keV, respectively.}
    \label{Fig:NuSTAR Energy resolved pulse profile}
\end{figure}

The spin period of the source was found to be 374.60(1) s for ObsID 02 using the epoch-folding tool \texttt{efsearch}. For ObsID 04, the spin period decreased to 374.62(1) s. The broadband (3$-$78 keV) pulse profiles from the two observations are shown in the top panel of Fig.~\ref{Fig:NuSTAR Energy resolved pulse profile}, where the minima of the two pulse profiles are aligned. The pulse profile for ObsID 02 was double-humped in nature, with some subpulse structures. The pulse profile changed significantly for ObsID 04 with the overall shape change of the individual peaks. We studied the energy-resolved pulse profile for both observations. We extracted pulse profiles in the energy ranges of 3$-$7, 7$-$10, 10$-$20, 20$-$40, 40$-$60, and 60$-$78 keV.

The pulse profile of ObsID 02 has a distinct main peak and a smaller peak separated by a pulse phase of $\sim$0.5. With an increase in the energy, the overall pulse shape remains the same, with the amplitude of modulation changing with energy.

The broadband 3$-$78 keV pulse profile for ObsID 04 exhibited notable modifications in pulse shape compared to ObsID 02. Instead of a main and secondary peak, a smaller peak emerged near the phase corresponding to the main peak in ObsID 02, coupled with a broader peak around the phase where the second peak was observed in ObsID 02. The overall pulse shape changed significantly with energy in ObsID 04. The low-energy, folded profiles show multiple peaks below $\sim$ 20 keV. However, above 20 keV, the pulse profile becomes double humped with a primary peak and a smaller peak, similar to that in ObsID 02.

\begin{figure}
\centering
\includegraphics[width=0.99\hsize]{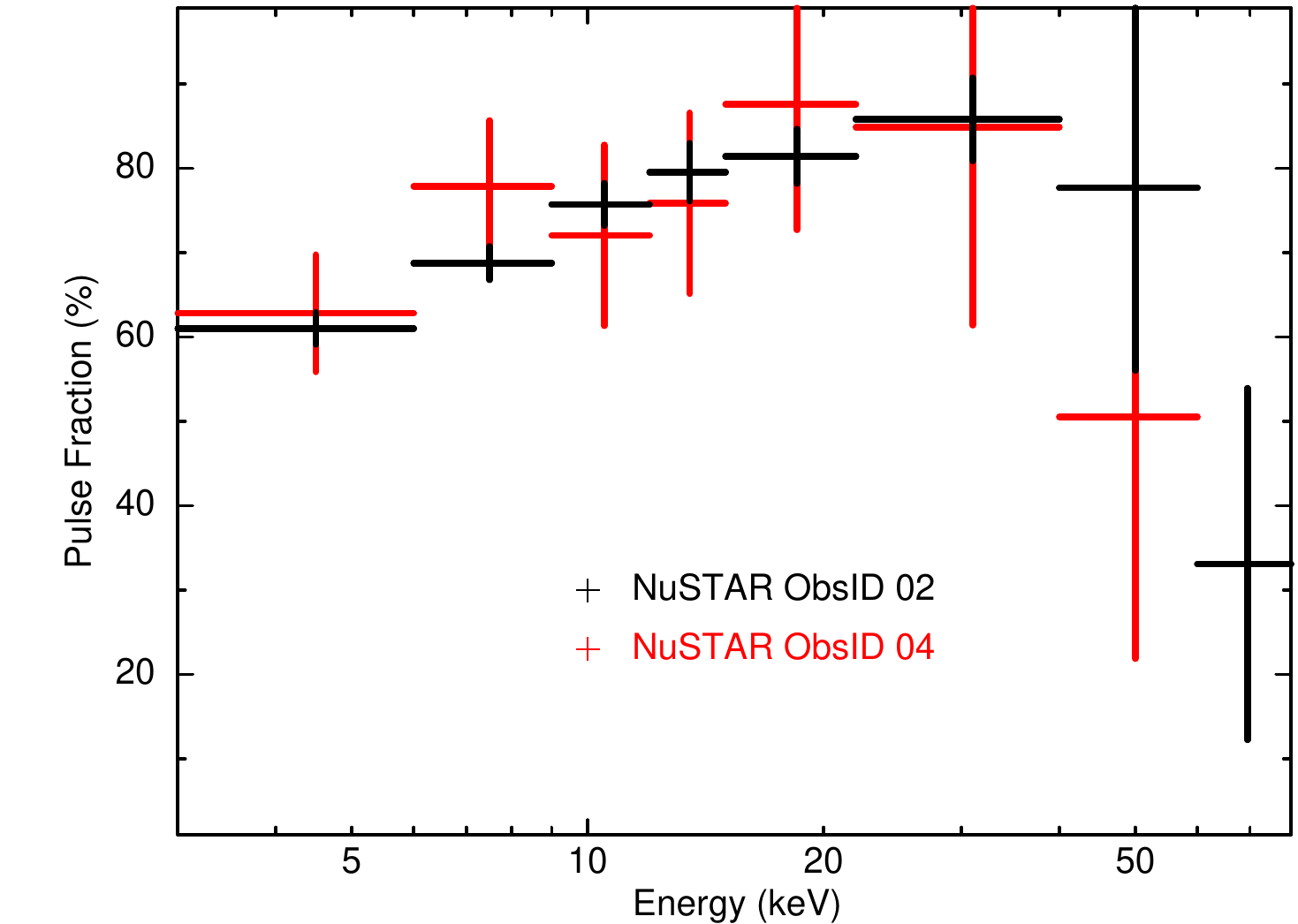}
    \caption{Variation of the pulsed fraction with energy for the two \textit{NuSTAR} observations ObsID 02 (\textit{in black}) and ObsID 04 (\textit{in red}). }
    \label{Fig:NuSTAR Pulse Fraction}
\end{figure}

To quantify the changes in modulation in the amplitude of pulse profiles, we define the pulse fraction (PF) as PF=(max-min)/(max+min), where max and min correspond to the maximum and minimum of the pulse profile. The variation of PF with energy is plotted in Fig.~\ref{Fig:NuSTAR Pulse Fraction} for both observations. We see an increasing trend of pulse fraction with energy until about 40 keV, beyond which there is a decrease in PF with energy, which is most prominent in ObsID 02. With the present data quality, it is not possible to comment on weather or not the decrease in pulse profile may be due to the presence of CRSF.

\subsection{Spectral analysis} \label{Spectral Analysis}

The phase-averaged spectra of IGR J0607.4+2205 from the two focal plane modules (FPMs) of \textit{NuSTAR} were fitted together allowing for a relative normalisation between the two \citep[in accordance with][]{FPM_crosscal_Madsen_2015}. All other parameters were tied between the two spectra. The entire spectral analysis was carried out using \texttt{XSPEC} version \texttt{12.12.1} \citep{XSPEC}.

The \textit{NuSTAR} broadband continuum spectrum of ObsID 02 was fitted with a power-law emission with a high-energy cutoff. The continuum model was multiplied by a \texttt{tbabs} spectral model to take into account the absorption of soft X-rays by interstellar matter with abundance taken from~\citet{Wilm_abund} and the photoelectric cross sections from~\citet{Vern_cs}. Residuals left in the form of absorption were fitted with a Gaussian absorption profile, and this could be interpreted as CRSF. The CRSF line was present at 51 $\pm$ 1 keV, with a large optical depth of $1.31_{-0.16}^{+0.14}$. Addition of the CRSF  improved the fit significantly and reduced the $\chi^2$(d.o.f.) from 880 (459) to 481 (456). This strong absorption feature was prominent in the raw spectrum as well. The residuals between spectral model and data  are shown in Fig.~\ref{Fig:NuSTAR
Obs 01 Spectrum}, with the cyclotron line optical depth set to zero. The significance of the CRSF feature was calculated by making 100,000 simulations of the data using the \texttt{XSPEC} tool \texttt{simftest}. In each of the  data simulations, spectral models were fitted with and without the absorption feature and the corresponding best-fit $\chi^2$ values were calculated. The false-detection probability was then calculated to be less than $10^{-5}$. The fluorescent iron K$\alpha$ line present in the spectrum was modelled using a \texttt{Gaussian} emission-line feature at $\sim$6.3 keV with an equivalent width of $16_{-4}^{+5}$ eV. This emission feature was quite significant with a false detection probability of $10^{-4}$ calculated from 10,000 simulations in the \texttt{simftest} tool. Power-law continuum with multiple cutoff models, such as \texttt{highEcut}, \texttt{highEcut} with a smoothing absorption feature\textbf{,} and \texttt{newhcut}~\citep{CenX3_Burderi_2000}, as well as \texttt{NPEX}~\citep{NPEX_Makishima_1999}, describe the continuum well. All models show residuals in high energy around 50 keV. The \texttt{newhcut} model was finally chosen as it yields the lowest $\chi^2$ in the fitting.

\begin{figure}
\centering
\includegraphics[width=0.99\hsize]{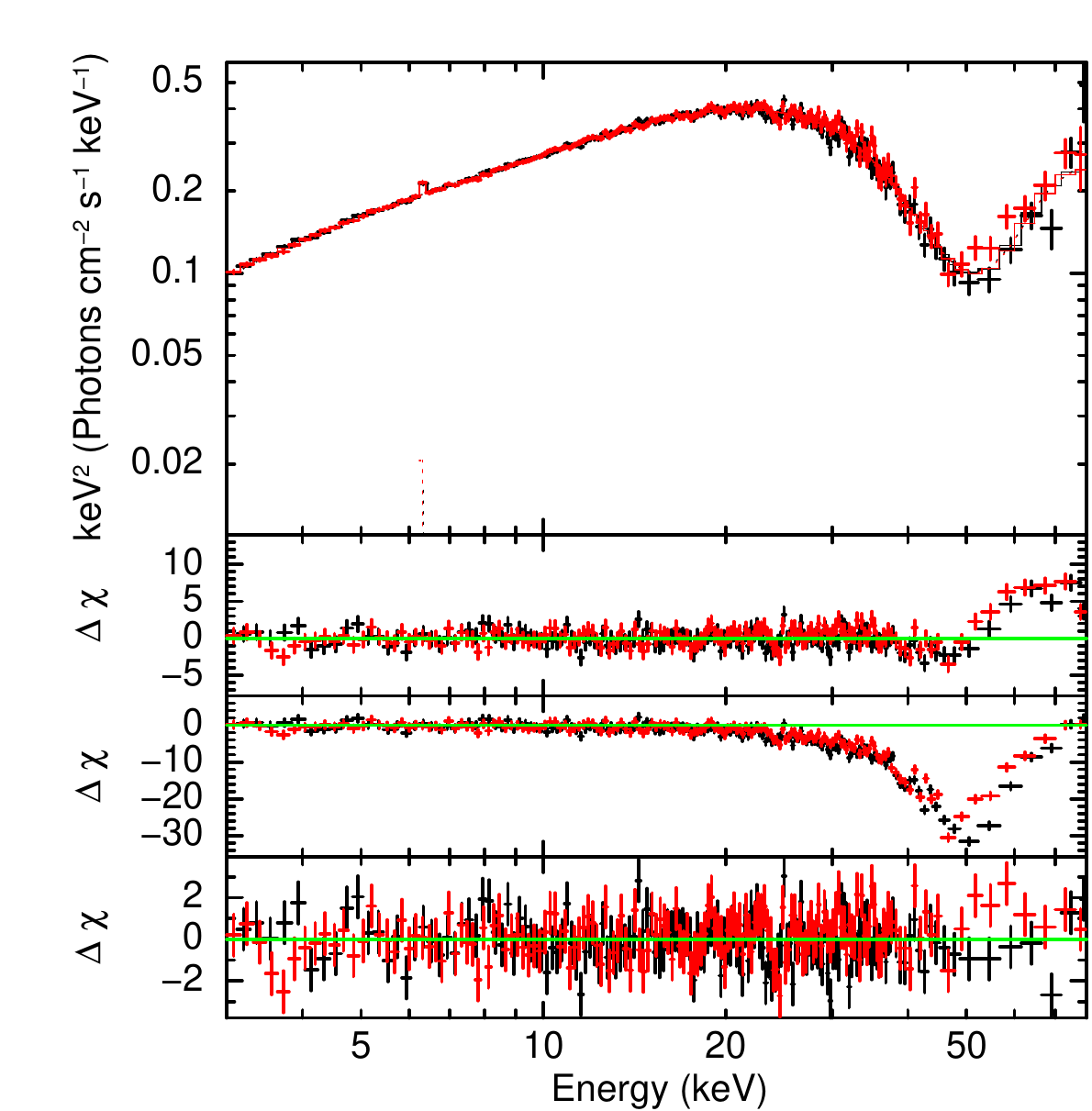}
    \caption{Spectrum and best-fit spectral model for ObsID 02. The top panel consists of the best-fit model along with the different components. The second panel from the top shows the residual from the best-fit model without the addition of a cyclotron line. The third panel from the top corresponds to the residual obtained by setting the strength of the CRSF to zero. The bottom panel corresponds to the best-fit model residuals. The black points correspond to FPMA, while the red points are for FPMB.}
    \label{Fig:NuSTAR Obs 01 Spectrum}
\end{figure}

The final spectral model used was  \texttt{tbabs} $\times$ ( \texttt{powerlaw} $\times$ \texttt{newhcut} + \texttt{Gaussian} ) $\times$ \texttt{gabs}. The power-law component has a photon index of 1.29$\pm$0.01, which is similar to those obtained from earlier \textit{XMM-Newton}~\citep{XMM_Newton_pulsation_Reig_2018} and \textit{Chandra}~\citep{BeXRB_Tomsick_2006ATel} observations. The best-fit spectral model in $E^2 dN/dE$ (top panel) as well as the residuals for the best-fit spectrum (bottom panel) are given in Fig.~\ref{Fig:NuSTAR Obs 01 Spectrum}. The best-fit spectral parameters are given in Table~\ref{Table:Spectral parameters} with errors at a 90\% confidence interval.

\begin{table}[]
\centering
\begin{tabular}{|c|c|c|}
\hline
Spectral Parameter                     & ObsID 02                & ObsID 04                \\ \hline
$N_H$ {[}$\times10^{22}cm^{-2}${]}     & $1.78_{-0.24}^{+0.27}$  & 0.43 (frozen)           \\ 
$\Gamma$                               & $1.29_{-0.01}^{+0.01}$  & $1.43_{-0.07}^{+0.06}$  \\ 
$\Gamma_{Norm}$ {[}$\times 10^{-2}${]} & $5.33_{-0.15}^{+0.17}$  & $0.41_{-0.01}^{+0.01}$  \\ 
$E_{cut}$ (keV)                        & $15.89_{-0.63}^{+0.65}$ & $30.93_{-1.95}^{+3.03}$ \\ 
$E_{fold}$ (keV)                       & $40.14_{-6.02}^{+7.00}$ & $11.81_{-2.89}^{+6.98}$ \\ 
Width                                  & $10.18_{-3.24}^{+2.57}$ & 5.0 (frozen)            \\ 
$kT_{BB}$ (keV)                        & $-$                     & $1.18_{-0.03}^{+0.03}$  \\ 
Radius$_{BB}$ (km)                     & $-$                     & $0.17_{-0.01}^{+0.01}$  \\ 
$E_{Fe}$ (keV)                         & $6.3_{-0.1}^{+0.1}$     & $-$                     \\ 
$EW_{Fe}$ (eV)                         & $16_{-4}^{+5}$          & $-$                     \\ 
$E_{CRSF}$ (keV)                       & $50.62_{-1.20}^{+0.97}$ & $-$                     \\ 
$\sigma_{CRSF}$ (keV)                  & $11.08_{-1.18}^{+0.93}$ & $-$                     \\ 
$\tau_{CRSF}$                          & $1.31_{-0.16}^{+0.14}$  & $-$                     \\ 
Cross Norm.                            & $0.99\pm0.01$  & $0.99\pm0.01$  \\ \hline
$\chi^{2}$ (d.o.f.)                    & 481 (456)               & 256 (242)               \\ 
Flux [$\times 10^{-10}$] $^a$          & $12.55_{-0.01}^{+0.01}$ & $0.86_{-0.11}^{+0.01}$  \\ 
Luminosity [$\times 10^{35}$] $^b$     & $30.42_{-0.34}^{+0.16}$ & $2.09_{-0.26}^{+0.02}$  \\ \hline
\end{tabular}
\caption{The table contains the best-fit spectral parameters of the IGR J0607.4+2205 for both observations. The errors are quoted at 90\% confidence intervals. $^{a}$ The flux is given in the energy range of 3$-$78 keV in units of erg cm$^{-2}$ s$^{-1}$. $^{b}$ The luminosity is given in the energy range of 3$-$78 keV in units of erg s$^{-1}$.}
\label{Table:Spectral parameters}
\end{table}

\begin{figure}
\centering
\includegraphics[width=0.95\hsize]{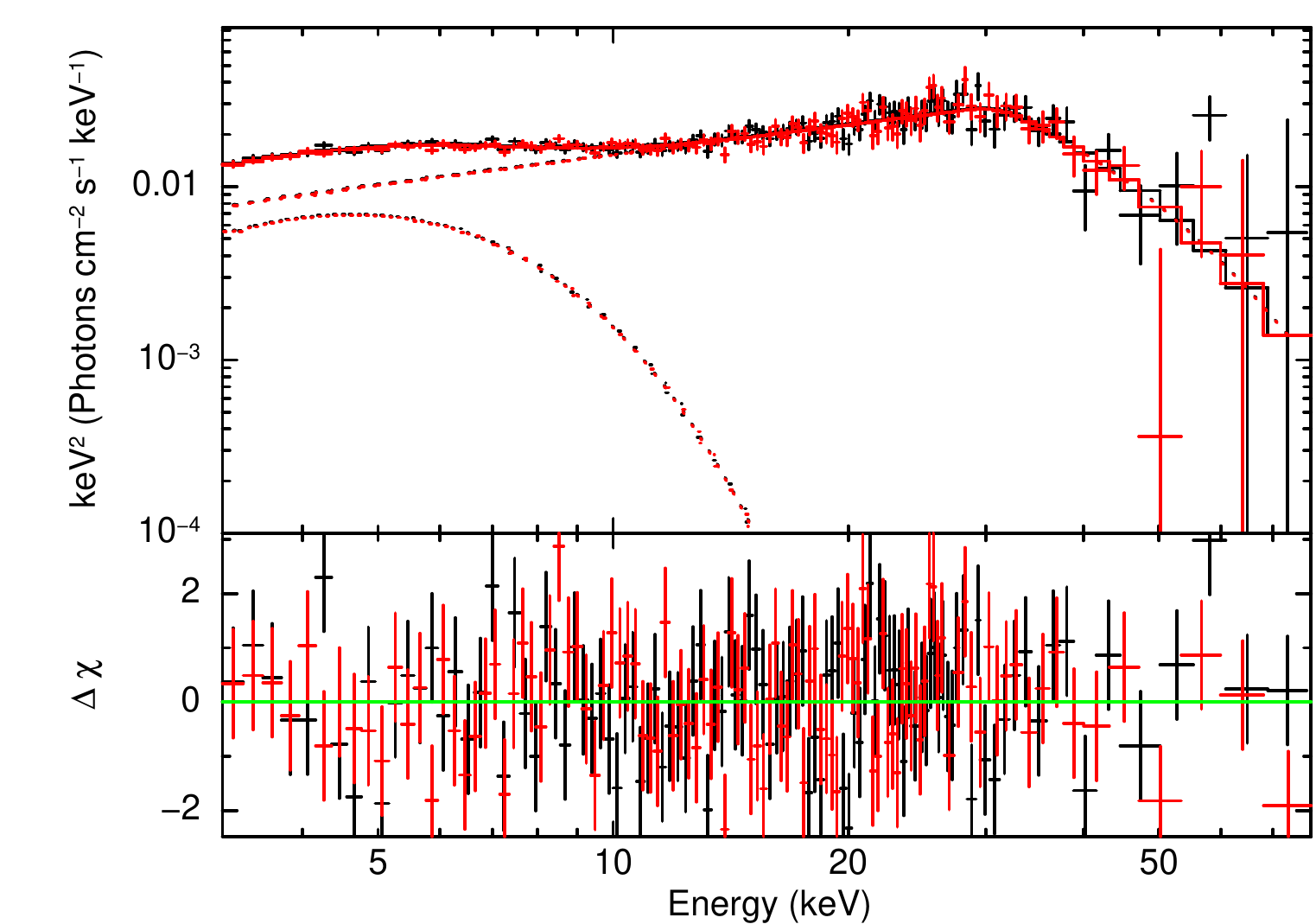}
    \caption{Spectrum and best-fit spectral model for ObsID 04. The top panel consists of the best-fit models along with the different components. The bottom panel corresponds to the best-fit model residuals. The black points correspond to FPMA, while the red points are for FPMB.}
    \label{Fig:NuSTAR Obs 02 Spectrum}
\end{figure}

A similar continuum model as of ObsID 02 was used for ObsID 04. There was a soft excess present in the spectrum of ObsID 04, which was modelled with a black body component. Addition of the black body component decreases the $\chi^2$(d.o.f.) from 446.58 (244) to 255.55 (242). The absorption column density cannot be constrained and was therefore fixed at a value of 0.43 $\times$ $10^{22}$ atoms cm$^{-2}$, which corresponds to the Galactic absorption component \textbf{along the line-of-sight}~\citep{Galactic_NH_Kalberla_2005, NH_HI4PI_Collaboration_2016}. The final spectral model of \texttt{tbabs} $\times$ ( \texttt{powerlaw} $\times$ \texttt{newhcut} + \texttt{bbodyrad} ) gave a $\chi^2$ of 256 for 242 degrees of freedom (d.o.f.). The temperature of the black body component was found to be 1.18 $\pm$ 0.03 keV, originating from an area of radius 0.17 $\pm$ 0.01 km. The high temperature and small radius of the thermal component indicate that it probably originates from the accretion mound near the magnetic poles of the NS. The best-fit spectra and residuals are given in Fig.~\ref{Fig:NuSTAR Obs 02 Spectrum}. In ObsID 04, the spectrum has very poor statistical significance above 40 keV, limiting the possibility for detection of a CRSF. As we are aware of the presence of a CRSF in the source for ObsID 02 at $\sim$ 51 keV, we added a Gaussian absorption component to the spectrum, with the line centre and width set to the best-fit values from ObsID 02. Nevertheless, the introduction of this extra component does not result in any enhancement in the $\chi^2$, yielding an upper limit on the optical depth of 0.68 at a 90$\%$ confidence level. The best-fit spectral parameters are provided in Table~\ref{Table:Spectral parameters}.

Subsequently, as we know there is a black body component from ObsID 04, we tried to add a similar black body component to ObsID 02. The temperature of the black body component cannot be constrained, and so it was frozen to the temperature from ObsID 04. The addition of the extra component leads to no significant decrease in $\chi^2$, yielding an upper limit on the luminosity of the black body component of $\sim10^{34}$ erg s$^{-1}$ at the 90$\%$ confidence level. The luminosity of the source was calculated assuming a distance of 4.5 kpc~\citep{Distance_Reig_2010}.

\section{Discussion}

IGR J0607.4+2205 underwent an outburst in October 2023. Two TOO observations were performed with \textit{NuSTAR,} capturing the source during an outburst  for the first time. In this work, we report the results from the two \textit{NuSTAR} observations of IGR J0607.4+2205 during the 2023 outburst of the source. 

Assuming a distance of 4.5 kpc~\citep{Distance_Reig_2010}, the source luminosity in the 3$-$78 keV range underwent a substantial change from $3.04 \times 10^{36}$ erg/s in ObsID 02 to $2.09 \times 10^{35}$ erg/s in ObsID 04. The period of pulsation obtained with \textit{NuSTAR} is consistent with the spin frequency observed in the \textit{XMM-Newton} observation, indicating no significant long-term slow down of the pulsar~\citep{XMM_Newton_pulsation_Reig_2018}. The persistent pulsations present during ObsID 04 suggest disc accretion of matter onto the NS. The pulse profile of IGR J0607.4+2205 is double-humped in nature with a high pulse fraction ($\geq$50$\%$). The source exhibits significant changes in pulse shape during the outburst. Luminosity-dependent changes in pulse profile have been observed in many X-ray pulsars~\citep[e.g.][]{EXO2030p375_Parmar_1989, V0332p53_Doroshenko_2017, 4U1626m67_Sharma_2023}. The low-energy ($\leq$20 keV) pulse shape of IGR J0607.4+2205 is significantly different between the two observations, while the higher energy (20$-$40 keV) pulse profile remains consistent. Similar behaviour has been reported for RX J0529.8$-$6556~\citep{Treiber_outburst_PP_change_2021}, where a misaligned Be disc is proposed to explain the observed changes in the pulse shape.

We modelled the broadband spectrum of the source with a cutoff power-law model as expected for HMXBs~\citep{HMXB_review_White_1983}. The hard X-ray spectrum of the source is dominated by a strong non-thermal (power-law) component with a high-energy cutoff during ObsID 02. We observe a CRSF for the first time in the spectrum of IGR J0607.4+2205, as expected from a highly magnetised BeXRP. The cyclotron line was detected at a very high energy of $50.6_{-1.2}^{+1.0}$ keV. From the CRSF centroid energy, we calculated the strength of the magnetic field to be 4.4$\pm$0.1 $\times$ $10^{12}$ G. We also report the first ever detection of iron K$\alpha$ fluorescent emission from IGR J0607.4+2205 with an equivalent width of $16_{-4}^{+5}$ eV. During ObsID 04, the source flux reduced by a factor of  about 15. A significant change in the spectral properties was observed and no CRSF feature can be constrained, which is possibly due to poor statistics. An additional thermal (black-body) component at a temperature $\sim$ 1 keV originating from a radius of $\sim$170 m was present during ObsID 04. The parameters of the black-body are consistent with thermal emissions from a hot spot on the surface of the NS.

Spectral parameters have been seen to evolve significantly in BeXRBs during outbursts \citep[e.g.][]{Reig2013, Lutovinov2021, Mandal2023}. The correlation between the spectral slope and the source luminosity has been found to change at a critical luminosity, signifying an alteration in the emission mechanism and beaming pattern of the source \citep{Becker2012}. From the CRSF \citet{Becker2012}, the critical luminosity of IGR J0607.4+2205 \textbf{was calculated} to be about $7 \times 10^{37}$ erg s$^{-1}$,  which is much higher than the observed luminosity during ObsID 02. This suggests that the source was well below the critical luminosity. The luminosity during ObsID 04 dropped to $\sim 2\times 10^{35}$ erg s$^{-1}$, where accretion onto the NS surface is near free fall and radiation is dominated by the hot spot or mounds near the NS polar caps \citep{Mushtukov2021}.

The discovery of a CRSF in a new source is crucial for determining the magnetic field strength near the NS surface and advancing our understanding of accretion processes near the surface of the NS. The accretion torque acting on an NS due to mass accretion depends on its mass-accretion rate, spin period, and magnetic field. Knowledge of the magnetic field strength is therefore crucial in order to corroborate the accretion torque models. Of the hundreds of X-ray binary pulsars, around 50 or so sources exhibit a CRSF~\citep{CRSF_review_Staubert_2019, CRSF_GROJ1750m27_AD}. The centroid energy of the CRSF detected in IGR J0607.4+2205 is relatively high among sources in which a CRSF has been detected. Additional broadband observations of IGR J0607.4+2205 are needed to study the luminosity dependence of the pulse profile and the CRSF.


\begin{acknowledgements}
      We thank the referee for the useful comments that improved the quality of this paper. KR would like to thank Ashwin Devaraj for his valuable inputs during the preparation of the manuscript for this project. This research made use of data obtained with \textit{NuSTAR}, a project led by Caltech, funded by NASA and managed by NASA/JPL.
\end{acknowledgements}

\bibliographystyle{aa}
\bibliography{bibliography}

\end{document}